\documentclass[10pt,aps,twocolumn,floats,floatfix,amssymb,prd,superscriptaddress,nofootinbib]{revtex4-2}

\usepackage{graphicx}
\usepackage{amsmath,amssymb}
\usepackage{amsfonts}
\usepackage{xspace} 
\usepackage[usenames]{color}
\usepackage{xcolor}
\usepackage{dcolumn}
\usepackage{bm}
\usepackage{mathrsfs}

\definecolor{Burgundy}{RGB}{144,0,32}
\usepackage[colorlinks=true,linkcolor=Burgundy,citecolor=Burgundy, filecolor=Burgundy, urlcolor=Burgundy]{hyperref}

\usepackage[all]{hypcap} 
\usepackage[utf8]{inputenc} 
\usepackage{multirow}
\usepackage{etoolbox}
\usepackage{tikz}
\usepackage[normalem]{ulem}
\usepackage{comment}
\usepackage{gensymb}

\newcommand{\be}{\begin{equation}}
\newcommand{\ee}{\end{equation}}
\newcommand{\bea}{\begin{eqnarray}}
\newcommand{\eea}{\end{eqnarray}}

\newcommand{\dd}{\mathrm{d}}

\begin{document}

\title{Non-minimal light-curvature couplings and black-hole imaging}

\author{Raúl Carballo-Rubio}
\email{raul.carballorubio@iaa.csic.es}
\affiliation{Instituto de Astrof\'isica de Andaluc\'ia (IAA-CSIC),
Glorieta de la Astronom\'ia, 18008 Granada, Spain}
\affiliation{Center of Gravity, Niels Bohr Institute, Blegdamsvej 17, 2100 Copenhagen, Denmark}

\author{Héloïse Delaporte}
\email{heloise.delaporte@uj.edu.pl}
\affiliation{Department of Science and Technology, University of the Faroe Islands, Vestara Bryggja 15, FO-100 Tórshavn, Faroe Islands}
\affiliation{Instytut Fizyki Teoretycznej, Uniwersytet Jagielloński,
Lojasiewicza 11, 30-348 Kraków, Poland}

\author{Astrid Eichhorn}
\email{eichhorn@thphys.uni-heidelberg.de}
\affiliation{Institut f\"ur Theoretische Physik, Universit\"at Heidelberg, Philosophenweg 16, 69120 Heidelberg, Germany}

\author{Pedro G. S. Fernandes}
\email{fernandes@thphys.uni-heidelberg.de}
\affiliation{Institut f\"ur Theoretische Physik, Universit\"at Heidelberg, Philosophenweg 16, 69120 Heidelberg, Germany}

\begin{abstract}
Non-minimal couplings between the electromagnetic field strength and the spacetime curvature are part of the effective field theory of gravity and matter. They alter the local propagation of light in a significant way if the ratio of spacetime curvature to the non-minimal coupling is of order one. Spacetime curvature can become appreciable around black holes, and yet the effect of non-minimal couplings on electromagnetic observations of black holes remains underexplored. A particular feature of the non-minimal coupling between the electromagnetic field-strength and the Riemann tensor is that it generates two distinct photon rings for different polarizations. Working within the paradigm of lensing bands and focusing on the $n=1$ lensing band, we illustrate by which diagnostics a modified light propagation may be distinguished from a modified spacetime geometry and how constraints on the value of the non-minimal coupling can be obtained.
\end{abstract}

\maketitle 

\section{Introduction}
Non-minimal interactions between light and spacetime curvature are  ubiquitous. For instance, they appear in the Euler-Heisenberg action on a curved background, i.e., they are generated in the low-energy regime of the Einstein-Maxwell theory  by integrating out quantum fluctuations of the electron \cite{Drummond:1979pp,Bastianelli:2008cu}. Even at a purely classical level, non-minimal interactions are allowed as part of the most general
Lagrangian for gravitational and electromagnetic fields that leads to second-order equations of motion~\cite{Horndeski:1976gi,BeltranJimenez:2013btb}.

The non-minimal interactions are built from various contractions of the electromagnetic field-strength $F_{\mu\nu}$ with the Riemann tensor $R_{\mu\nu\kappa\lambda}$ and its own contractions, as well as with the Levi-Civita tensor $\epsilon_{\mu\nu\rho \sigma}$. Possible interactions include, e.g., $F_{\mu\nu}F_{\rho\sigma}C^{\mu\nu\rho\sigma}$ with the Weyl tensor  $C^{\mu\nu\rho\sigma}$, which was recently investigated in quantum gravity \cite{Knorr:2024yiu}. Another example is the Horndeski term, which couples the double-dual of the Riemann tensor to two powers of the  electromagnetic field-strength~\cite{Horndeski:1976gi,BeltranJimenez:2013btb}.
 
The lowest-order non-minimal interactions are dimension-six operators and, in accordance with standard effective-field-theory (EFT) arguments, are expected to be suppressed by the square of the mass-scale of the particle that is integrated out, e.g., the electron. If a very light and milli-charged particle exists\footnote{See \cite{Davidson:2000hf,Boyarsky:2018tvu,Alvey:2020xsk,deMontigny:2023qft,Kalliokoski:2023cgw} for up-to-date studies of the parameter space.}, then the corresponding mass scale can be much lower, rendering those operators less suppressed
at lower energies. From a different perspective, positivity bounds  constrain such couplings~\cite{Bellazzini:2019xts}, such that the EFT does not allow arbitrary coefficients.

It is also expected that quantum gravitational fluctuations generate such terms. Asymptotically safe quantum gravity is an explicit example of a candidate theory
that gives rise to such terms \cite{Knorr:2024yiu}. Also within a quantum-gravitational context, the couplings between light and curvature are restricted by the weak-gravity conjecture \cite{Bellazzini:2019xts,Knorr:2024yiu}.  

From a more phenomenological perspective, non-minimal couplings could be relevant in settings in which light propagates on high-curvature backgrounds~\cite{Drummond:1979pp,Daniels:1993yi,Daniels:1995yw,Shore:1995fz,Shore:2002gn,Jana:2021lqe}.  Important examples are black-hole images, which arise when light propagates on a black-hole spacetime. Following (and in part even preceding) the first images of M87* 
\cite{EventHorizonTelescope:2019dse,EventHorizonTelescope:2019uob,EventHorizonTelescope:2019jan,EventHorizonTelescope:2019ths,EventHorizonTelescope:2019pgp,EventHorizonTelescope:2019ggy,EventHorizonTelescope:2021bee,EventHorizonTelescope:2021dqv,EventHorizonTelescope:2021srq,EventHorizonTelescope:2023gtd} and Sgr A* \cite{EventHorizonTelescope:2022wkp,EventHorizonTelescope:2022apq,EventHorizonTelescope:2022wok,EventHorizonTelescope:2022exc,EventHorizonTelescope:2022urf,EventHorizonTelescope:2022xqj,EventHorizonTelescope:2024hpu,EventHorizonTelescope:2024rju} and their subsequent observations \cite{EventHorizonTelescope:2024dhe,EventHorizonTelescope:2024uoo}, numerous studies explore the potential to constrain deviations from General Relativity (GR), see, e.g.,~\cite{Broderick:2013rlq,Wei:2013kza,Grenzebach:2014fha,Johannsen:2015hib,Cunha:2015yba,Cunha:2016wzk,Vincent:2016sjq,Abdujabbarov:2016hnw,Amir:2016cen,Tsukamoto:2017fxq,Ayzenberg:2018jip,Psaltis:2018xkc,Mizuno:2018lxz,Cunha:2018acu,Held:2019xde,Tian:2019yhn,Vagnozzi:2019apd,Allahyari:2019jqz,Cunha:2019dwb,Volkel:2020xlc,Konoplya:2020bxa,Khodadi:2021gbc,Younsi:2021dxe,EventHorizonTelescope:2022xqj,Vagnozzi:2022moj,Eichhorn:2022fcl,Carballo-Rubio:2022aed,Chen:2022scf,Khodadi:2020jij,Fernandes:2024ztk,dePaula:2023ozi,Sengo:2022jif,Junior:2021dyw,Xavier:2020egv,Olmo:2023lil,Ayzenberg:2023hfw,Carballo-Rubio:2023ekp,Afrin:2022ztr,Salehi:2023eqy,Lupsasca:2024wkp,Broderick:2024vjp,Khodadi:2024ubi,Carballo-Rubio:2025fnc,Bambi:2025wjx} and references therein. 
In these studies, it is typically assumed that the parameters that control the deviation from GR are large, i.e., the dimensionless ratio of beyond-GR couplings to an appropriate power of the gravitational radius is typically assumed to be $\mathcal{O}(1)$. From an EFT perspective and invoking a naturalness argument, such a setting should also feature non-minimal couplings.
However, effects that arise from non-minimal couplings have been significantly less studied~\cite{Chen:2015cpa,Chen:2016hil,Jana:2021lqe,Chen:2023wna}. In building a physics-case for future Very Long Baseline Interferometry (VLBI)-arrays, either ground-based \cite{Doeleman:2023kzg,Johnson:2023ynn, Ayzenberg:2023hfw} or space-based \cite{2024SPIE13092E..6PI,Johnson:2024ttr,Zineb:2024gwx}, it is important to understand which new-physics effects can (i) be tested in principle and (ii) be distinguished from each other (see, e.g.,~\cite{Glampedakis:2021oie,Lara:2021zth,Ozel:2021ayr,Kocherlakota:2022jnz}). For this goal, it is important to figure out whether non-minimal  couplings, which modify the propagation of light  on a curved spacetime, can have similar effects to the modification of the spacetime itself. With this motivation, we explore the impact of a non-minimal photon-curvature coupling on black-hole image features.

This paper is structured as follows: In Sec.~\ref{sec:theory} we review the Lagrangian that we will study. In Sec.~\ref{sec:lightprop} we discuss the propagation of light in this setting, before turning to black-hole lensing bands in Sec.~\ref{sec:lensingbands}. We present our results on lensing bands and ranges of the coupling that can be excluded given a photon-ring observation in Sec.~\ref{sec:results}, before concluding in Sec.~\ref{sec:conclusions}.

\section{Non-minimal light-curvature couplings and effective field theory}\label{sec:theory}

Non-minimal couplings of the electromagnetic field-strength $F_{\mu\nu}$ to the spacetime curvature without negative powers of derivatives are necessarily at least dimension-six operators and  contain four derivatives, because otherwise $U(1)$ gauge symmetry and/or diffeomorphism invariance cannot be respected. Models with more than two time derivatives are difficult to treat, because, according to the Ostrogradsky theorem from classical mechanics, such field theories have a Hamiltonian that is unbounded from below under a non-degeneracy assumption.\footnote{It is usually assumed that such an unbounded Hamiltonian results in a catastrophic instability of the theory. Recently, this assumption has been challenged by explicit counterexamples as well as proofs of well-defined time-evolution \cite{Deffayet:2021nnt,Deffayet:2023wdg,Deffayet:2025lnj,Held:2025ckb}. A full theory of when higher-order time derivatives do not lead to catastrophic instabilities has not yet been reached. Once it has become available, we expect that a larger subset of EFTs for photon-curvature interactions will become accessible.}
Thus, we focus on the unique non-minimal interaction that can be added to the Einstein-Maxwell Lagrangian without resulting in higher-order equations of motion. This interaction was derived by Horndeski \cite{Horndeski:1976gi} and leads to the Horndeski vector-tensor theory given by the Lagrangian \cite{Horndeski:1976gi,BeltranJimenez:2013btb}
\begin{equation}\label{eq:vth_action}
    \mathcal{L} = \frac{M_{\rm pl}^2}{2}R - \frac{1}{4} F_{\mu \nu}F^{\mu \nu} + \alpha L_{\mu \nu \rho \sigma}F^{\mu \nu} F^{\rho \sigma},
\end{equation}
where $F_{\mu\nu}=\nabla_\mu A_\nu-\nabla_\nu A_\mu$,
\begin{equation}
    L^{\mu\nu\rho\sigma}=-\frac{1}{4}\epsilon^{\mu\nu\alpha\beta}\epsilon^{\rho\sigma\gamma\delta}R_{\alpha\beta\gamma\delta}
\end{equation}
is the double-dual Riemann tensor, $\epsilon_{\mu\nu\alpha\beta}$ the Levi-Civita tensor and $\alpha$ is a coupling constant with dimensions of length squared\footnote{We work in units in which $\hbar = 1= c$.} controlling the strength of the non-minimal coupling to gravity.\footnote{Although the strength and sign of $\alpha$ are arbitrary, they might be constrained by stability requirements. For an exterior Schwarzschild background to be stable, $\alpha$ can take any sign, but must be subdominant compared to the minimal coupling term, see \cite{BeltranJimenez:2013btb}.} The Horndeski term can also be expressed as
\begin{eqnarray}
        L_{\mu \nu \rho \sigma}F^{\mu \nu} F^{\rho \sigma} &=& R F_{\mu \nu}F^{\mu \nu} -4 R_{\mu \nu} F^{\mu \rho} F^{\nu}_{\phantom{\nu} \rho} \nonumber\\
        &{}&+ R_{\mu \nu \rho \sigma} F^{\mu \nu}F^{\rho \sigma}.
\end{eqnarray}
The Horndeski vector-tensor term also emerges in the effective four-dimensional theory obtained by a Kaluza-Klein reduction of higher-dimensional Gauss-Bonnet gravity \cite{Buchdahl:1979wi}.

The above Lagrangian should not be understood as a proposal for a fundamental theory. It may rather be understood as an EFT-description of the gravity-vector system in which the coupling $\alpha$ is parametrically larger than the other couplings in the fourth order of the derivative expansion.\footnote{Often in EFT treatments, it is assumed that dimensionless couplings are of $\mathcal{O}(1)$, such that all dimensionful couplings are set by the same mass scale. This need not in general be the case, because the microscopic theory may contain mechanisms to selectively enhance just a subset of the couplings in the EFT compared to the other couplings at the same order of the derivative expansion. One example of such a mechanism relies on non-trivial scaling dimensions which arise due to quantum fluctuations, see, e.g., \cite{Brenner:2024bps} for a specific example.
Thus, the above Lagrangian may even be a complete EFT (to that order).} However, it is best understood as a toy model, in which we can explore which types of effects may arise due to non-minimal interactions and we choose the model guided by technical simplicity, in particular requiring second-order equations of motion. 
As we will see, the model is already complex enough to exhibit several distinct effects in black-hole images.

Previously, Ref.~\cite{Prasanna:2003ix} derived bounds on the non-minimal coupling $\alpha$. From the birefringent bending of light by the sun, Ref.~\cite{Prasanna:2003ix} obtained $\sqrt{|\alpha|} \lesssim 3.16\times 10^{5} \, \mathrm{km}$, while from radar echo experiments of radio signals passing in the vicinity of the sun they obtained $\sqrt{|\alpha|} \lesssim 6.25\times 10^{4} \, \mathrm{km}$, and $\sqrt{|\alpha|} \lesssim 2.45 \, \mathrm{km}$ from the timing of the binary pulsar PSR B1534+12 signals. Taking these constraints at face value would rule out $|\alpha|/M^2\sim \mathcal{O}(1)$ (for $M$ the mass of a supermassive black hole), which is the range of values for which we expect black-hole images to change appreciably. These bounds should in our view be interpreted with a grain of salt for the following reason. Astrophysical constraints on couplings are typically subject to a lack of control over the system (in contrast to laboratory experiments), and the present constraints are no exception. For instance, in some astrophysical environments, both magnetic fields as well as a non-vanishing Ricci scalar are present. Both affect the propagation of light, if higher-order field-strength terms as well as a non-minimal Ricci coupling are included in the Lagrangian. Translating observations into constraints is subject to assumptions about whether or not such couplings are present in Lagrangians. Existing constraints from astrophysical systems are derived under simplifying assumptions, such as the absence of a standard Euler-Heisenberg term (higher-order field strength term), which modifies the propagation of light in the presence of magnetic fields.

However, our main motivation to consider $\alpha$ is not that we are interested in observational constraints on this specific coupling. Rather, we consider $\alpha$ as one representative of a much larger class of non-minimal light-curvature couplings, for which it has never been explored before what their effects on photon rings are. It is, however, crucial to understand their effects, because, if future observational data on photon rings is used to constrain new physics in gravity, it is not sufficient to only consider modifications of the spacetime geometry. Accounting for new-physics effects that change the propagation of light is also crucial. Here, we take a first step in this direction. In doing so, we focus on $\alpha$ specifically, rather than other (so far unconstrained) non-minimal couplings, purely for reasons of calculational convenience, because $\alpha$, being part of the Horndeski family and having an effective metric description for photon propagation, c.f.~Sec.~\ref{sec:lightprop}, allows us to compute effects easily. We expect that the effects induced by $\alpha$ are paradigmatic of those created by non-minimal couplings more generally, motivating our study of their imprints on black-hole images.

\section{\protect Propagation of non-minimally
 coupled
photons
}\label{sec:lightprop}
 
In the presence of a non-minimal  coupling, the propagation of photons is, in general, not described by the standard geodesic equation~\cite{Drummond:1979pp}. The latter is obtained from the wave-like solutions of motion for minimally coupled electromagnetic fields in the geometric-optics limit. The equation of motion is modified by non-minimal interactions; therefore, the null geodesics of the background spacetime $g_{\mu\nu}$ lose their interpretation as describing the propagation of photons.

At a technical level, some settings with non-minimal couplings enable a description through the geodesic equation of an effective metric $\tilde{g}_{\mu\nu}$. This effective metric does not encode the geometry of spacetime; instead, it captures the effect of the non-minimal interaction at an effective level.
This can be done if there is a field transformation for the particular solution $g_{\mu\nu} \rightarrow \tilde{g}_{\mu\nu}$ that maps the Lagrangian in Eq.~\eqref{eq:vth_action} into the Einstein-Maxwell Lagrangian plus higher-curvature corrections. However, as we will see below, the situation for black-hole backgrounds is more complex and, in particular, a single effective metric does not suffice.

We focus on the Schwarzschild spacetime
\begin{eqnarray}
    g_{\mu\nu}\dd x^\mu\dd x^\nu &=& - f(r) \dd t^2 + f(r)^{-1} \dd r^2 + r^2\dd\Omega^2,
    \label{eq:sssmetric1}\\
   f(r) &=&  1 - \frac{2M}{r}\nonumber,
\end{eqnarray}
where $\dd\Omega^2= \dd \theta^2 + \sin^2\theta\, \dd \varphi^2$ is the standard metric of the unit 2-sphere. The Schwarzschild spacetime, which is Ricci flat ($R_{\mu \nu} = 0$), is a solution of the Horndeski vector-tensor theory. The vector part of the Horndeski vector-tensor theory on this background reduces to
\begin{equation}\label{eq:vth_action_riem}
    \mathcal{L}_{\rm em} = - \frac{1}{4} F_{\mu \nu}F^{\mu \nu} + \alpha R_{\mu \nu \rho \sigma}F^{\mu \nu} F^{\rho \sigma},
\end{equation}
since $L_{\mu \nu \rho \sigma} = R_{\mu \nu \rho \sigma}$, when $R_{\mu \nu}=0$.
This theory is also the minimal one we can consider that captures the effects of a non-minimal light-curvature coupling. The modified Maxwell equations read 
\begin{equation}\label{eq:maxeq}
    \nabla_{\rho} F^{\rho \mu} - 4 \alpha \nabla_{\rho}\left(R^{\mu\rho\sigma\nu} F_{\sigma\nu}\right) = 0.
\end{equation}
This equation can be analyzed in the geometric optics approximation, using the parametrization $F_{\mu\nu}=f_{\mu\nu}e^{i\theta}$, and assuming that derivatives of $f_{\mu\nu}$ are subleading with respect to the phase gradient $k_\mu=\partial_\mu\theta$. Contracting $k^\mu$ with the Bianchi identity $k_\mu f_{\nu\rho}+k_\nu f_{\rho\mu}+k_\rho f_{\mu\nu}=0$ for the electromagnetic field-strength, and using Eq.~\eqref{eq:maxeq}, we can write
\begin{equation}\label{eq:maxeqjk}
\left(k^2g^{\mu\sigma}g^{\nu\gamma} +8\alpha k^{\left[\mu\right.} R^{\left.\nu\right]\rho\sigma\gamma}k_\rho\right)f_{\sigma\gamma}=0, 
\end{equation}
where the square brackets correspond to the anti-symmetrization of indices as $T_{[\mu\nu]} =  (T_{\mu\nu} - T_{\nu\mu})/2$.
Eq.~\eqref{eq:maxeqjk} is a homogeneous system of equations for the tensor $f_{\mu\nu}$ that admits non-trivial solutions if and only if $k^\mu$ is suitably constrained. To extract the form of the relevant constraints, we note that the Riemann tensor for the Schwarzschild spacetime can be written as~\cite{Drummond:1979pp}
\begin{equation}\label{eq:schw_ansatz2}
R^{\mu\nu\rho\sigma}=\frac{M}{r^3}\left[g^{\mu\rho}g^{\nu\sigma}-g^{\mu\sigma}g^{\nu\rho}-3\sum_{i=1}^2(-1)^iU_{(i)}^{\mu\nu}U_{(i)}^{\rho\sigma}\right],
\end{equation}
where $\{U_{(i)}^{\mu\nu}\}_{i=1}^2$ are defined in terms of vierbein fields $\{e^\mu_{(a)}\}_{a=0}^3$ satisfying $g^{\mu\nu}=\eta^{ab}e^\mu_{(a)}e^\mu_{(b)}$, with $\eta^{ab}$ the inverse flat metric, as~\cite{Drummond:1979pp}
\begin{align}
U_{(1)}^{\mu\nu}&=e^\mu_{(0)}e^\nu_{(1)}-e^\mu_{(1)}e^\nu_{(0)},\nonumber\\
U_{(2)}^{\mu\nu}&=e^\mu_{(2)}e^\nu_{(3)}-e^\mu_{(3)}e^\nu_{(2)}.
\end{align}

Inserting this expression of the Riemann tensor into Eq.~\eqref{eq:maxeqjk}, the constraints that must be satisfied by $k^\mu$ take the form
\begin{equation}
\tilde{g}^{(\bm{v})}_{\mu\nu}k^\mu k^\nu = 0,
\end{equation}
with effective metrics\footnote{Note that these effective metrics do not encode the actual spacetime geometry, thus indices are always raised and lowered with $g_{\mu\nu}$, not $\tilde{g}_{\mu\nu}^{(\bm{v})}$.} $\tilde{g}_{\mu\nu}^{(\bm{v})}$
depending on the following two independent polarization vectors $l^\nu$ and $m^\nu$
\begin{equation}
    \begin{aligned}
        &l^\nu=k_\mu U^{\mu\nu}_{(1)} =(-k_1,k_0,0,0),\\&
        m^\nu=k_\mu U^{\mu\nu}_{(2)}=(0,0,-k_3,k_2)/r^2\sin\theta.
    \end{aligned}
    \label{eq:pol_vectors}
\end{equation}
It holds that $k^{\mu}l_{\mu}=0=k^{\mu}m_{\mu}$, so that $l_{\mu}$ and $m_{\mu}$ can be understood as the two real basis vectors which encode the linear polarization of a photon in the plane orthogonal to its direction of propagation. These orthogonal linear polarizations are chosen based on theoretical considerations, as they encapsulate the effects of the effective metrics induced by the non-minimal light-curvature coupling. However, linearly polarized radiation, as measured, for M87*, by the EHT Collaboration \cite{EventHorizonTelescope:2021bee}, is usually expressed as a combination of the three real Stokes parameters $\mathcal{I, Q, U}$ encoding the total intensity of the radiation and its linearly polarized components, see \cite{Thompson:2017int}. While a mapping exists between the two descriptions, we will solely focus on the theoretical one provided by $l_{\mu}$ and $m_{\mu}$ in Eq.~\eqref{eq:pol_vectors}, as we will not compare our results with actual observational data on polarization. 
The corresponding effective metrics can be written (up to an arbitrary conformal factor, because the effective metrics only encode modified propagation of light) as~\cite{Chen:2015cpa}
\begin{equation}
    \tilde{g}^{(\bm{v})}_{\mu\nu}\dd x^\mu\dd x^\nu = -f(r) \dd t^2 + f(r)^{-1}\dd r^2 + \rho^{(\bm{v})}(r) r^2 \dd\Omega^2,
    \label{eq:sssmetric2}
\end{equation}
where $\bm{v}$ is either $\bm{l}$ or $\bm{m}$, and
\begin{equation}
    \rho^{(\bm{l})}(r) = \frac{r^3 - 8 \alpha M}{r^3 + 16 \alpha M},
\end{equation}
for a photon with polarization along $l^\nu$ (in short, PPL), and
\begin{equation}
    \rho^{(\bm{m})}(r) = \frac{r^3 + 16 \alpha M}{r^3 - 8 \alpha M},
\end{equation}
for a photon with polarization along $m^\nu$ (in short, PPM). 
Within the geometric optics approximation used to obtain the effective metric for light propagation, this limits the coupling to
\begin{equation}
    -1/2 \leq \alpha/M^2 \leq 1,
    \label{eq:couplingrange}
\end{equation}
in order for the effective metrics to be non-singular at the horizon $r=2M$ for both polarizations. Note that the radial coordinates of the metrics in Eqs.~\eqref{eq:sssmetric1} and \eqref{eq:sssmetric2} have different physical interpretations. While in Eq.~\eqref{eq:sssmetric1}, $r$ corresponds to the geometrically meaningful areal radius, this is not the case for Eq.~\eqref{eq:sssmetric2}, whose areal radius is given by $r_{\rm areal} = r\sqrt{\rho^{(\bm{v})}(r)}$. Consequently, although the horizon is located at $r=2M$ in both cases, the areal radius of the horizon for photons in the effective metrics is $r_{\rm areal,H} = 2M\sqrt{\rho^{(\bm{v})}(2M)}$.

In practice, the effective metrics above indicate that light rays do not propagate along null geodesics of the metric $g_{\mu\nu}$. Photons with different polarizations propagating on the Schwarzschild spacetime experience different effective metrics in the geometric optics approximation (later studies also considered the cases of charged and rotating black-hole spacetimes \cite{Daniels:1993yi,Daniels:1995yw} and obtained the effective metrics in the geometric optics approximation, see, e.g.,~\cite{Chen:2016hil}). This is a phenomenon known as bi-metricity~\cite{Visser:2002sf}. In the sections below, we will determine the impact this feature can have on black-hole images.

\section{Calculation of lensing bands\label{sec:lensingbands}} 
\subsection{Background on lensing bands}
To explore  how
the non-minimal coupling $\alpha$  affects VLBI observations of supermassive black holes, we leverage the concept of lensing bands. Lensing bands have been explored in \cite{Paugnat:2022qzy, Cardenas-Avendano:2023obg} as powerful probes of new physics beyond GR, because they marginalize over astrophysical uncertainties. 

More specifically, a black-hole image always depends on (i) the spacetime geometry, (ii) the propagation equation for light and (iii) the distribution and properties of matter that emits and absorbs light. We are interested in testing (i) and (ii) independently of (iii), because our understanding of the astrophysical properties of accretion disks is not well enough developed to uniquely specify the emission and absorption profile of the accretion disk of a given black hole \cite{Lara:2021zth,Ozel:2021ayr,Kocherlakota:2022jnz,EventHorizonTelescope:2022urf,Chatterjee:2022pbo}. In addition, accretion disks are not always in a stationary state, so that time-dependence of the emission is in general expected \cite{EventHorizonTelescope:2022urf,Murchikova:2022aiz,Chatterjee:2022pbo,Blaes:2025usm}.

There are two ways of addressing this uncertainty, both of which rely on photon rings. Photon rings arise because of strong gravitational lensing around a black hole. They constitute lensed images of the black hole which are present in addition to the direct image and which are exponentially stacked in GR.
The $n$th photon ring in a black-hole image is constituted by those geodesics that, roughly speaking, complete $n$ half-orbits about the black hole, see \cite{Eichhorn:2022bbn} for a discussion of how to define a half-orbit in and beyond GR and \cite{Johnson:2019ljv,Gralla:2019xty,Gralla:2020srx,Gralla:2020yvo,Broderick:2021ohx,Chael:2021rjo} for a discussion of photon rings. It can be shown that the direct emission that arrives on an observer's screen from the vicinity of a black hole (the $n=0$ emission) depends on where in the spacetime the light was emitted, i.e., the location of the $n=0$ emission in the image is strongly dependent on the astrophysical properties of the accretion disk. For $n=1$, the dependence is already lower,  as light that completes  one half-orbit about the black hole before arriving at the observer's screen necessarily arrives within a bounded region in the image, known as the $n=1$  photon ring. As $n$ increases, the size of this bounded region decreases and is therefore less and less dependent on the properties of the accretion disk. For $n \rightarrow \infty$, no dependence on the properties of the accretion disk is left. Therefore, photon rings constitute observables that probe spacetime geometry largely independently of astrophysical properties of the accretion disk, see, e.g., \cite{Broderick:2021ohx,Wielgus:2021peu,Eichhorn:2022oma,Ayzenberg:2022twz,Staelens:2023jgr,Kocherlakota:2023qgo,Aratore:2024bro,Galison:2024bop,KumarWalia:2024yxn,Farah:2024mkq,KumarWalia:2024omf,Keeble:2025gbj}.

However, the $n \rightarrow \infty$ photon ring is not observable, because it has vanishing intensity. In practice, the $n=1$ photon ring may be detectable by future VLBI-arrays, see, e.g., \cite{Ayzenberg:2023hfw,Lupsasca:2024xhq,Galison:2024bop,Tamar:2024tfc}. 
Imposing an informed prior on the image reconstruction for M87*, it has been shown in \cite{Broderick:2022tfu} that a persistent (across consecutive observational days) thin ring is statistically favored in addition to a broad emission region over reconstructed images without a thin ring. Thin photon rings are generic in GR and beyond, thus motivating this choice of prior. This thin ring may be interpreted as the $n=1$ photon ring, but its reconstruction is of course predicated on the choice of prior.

The location of the $n=1$ photon ring in the image depends on all three aspects (i-iii), i.e.~the spacetime and light propagation properties as well as the properties of the accretion disk. However, the $n=1$ photon ring must be contained within a bounded region in the image, known as the $n=1$ lensing band. The $n=1$ lensing band is defined as the set of points on the observer's screen to which at least one photon completing one half-orbit about the black hole arrives after being emitted at an arbitrary location on the equatorial plane.  Accordingly, the lensing band can be understood as the region in the image within which the location of the $n=1$ photon ring varies when the emission region is varied in-between the black-hole horizon and infinitely far away. By definition, the $n=1$ photon ring always lies within the $n=1$ lensing band.

As highlighted in \cite{Cardenas-Avendano:2023obg}, if the $n=1$ lensing band in a theory beyond GR does not overlap with the corresponding lensing band in GR for a given range of couplings, an $n=1$ photon ring detection would either rule out this range of couplings or rule out GR, with a confidence level that depends on the uncertainties in the photon ring detection. 

\subsection{Numerical method}
\begin{figure*}[t!]
    \centering
    \includegraphics[width=0.333\linewidth]{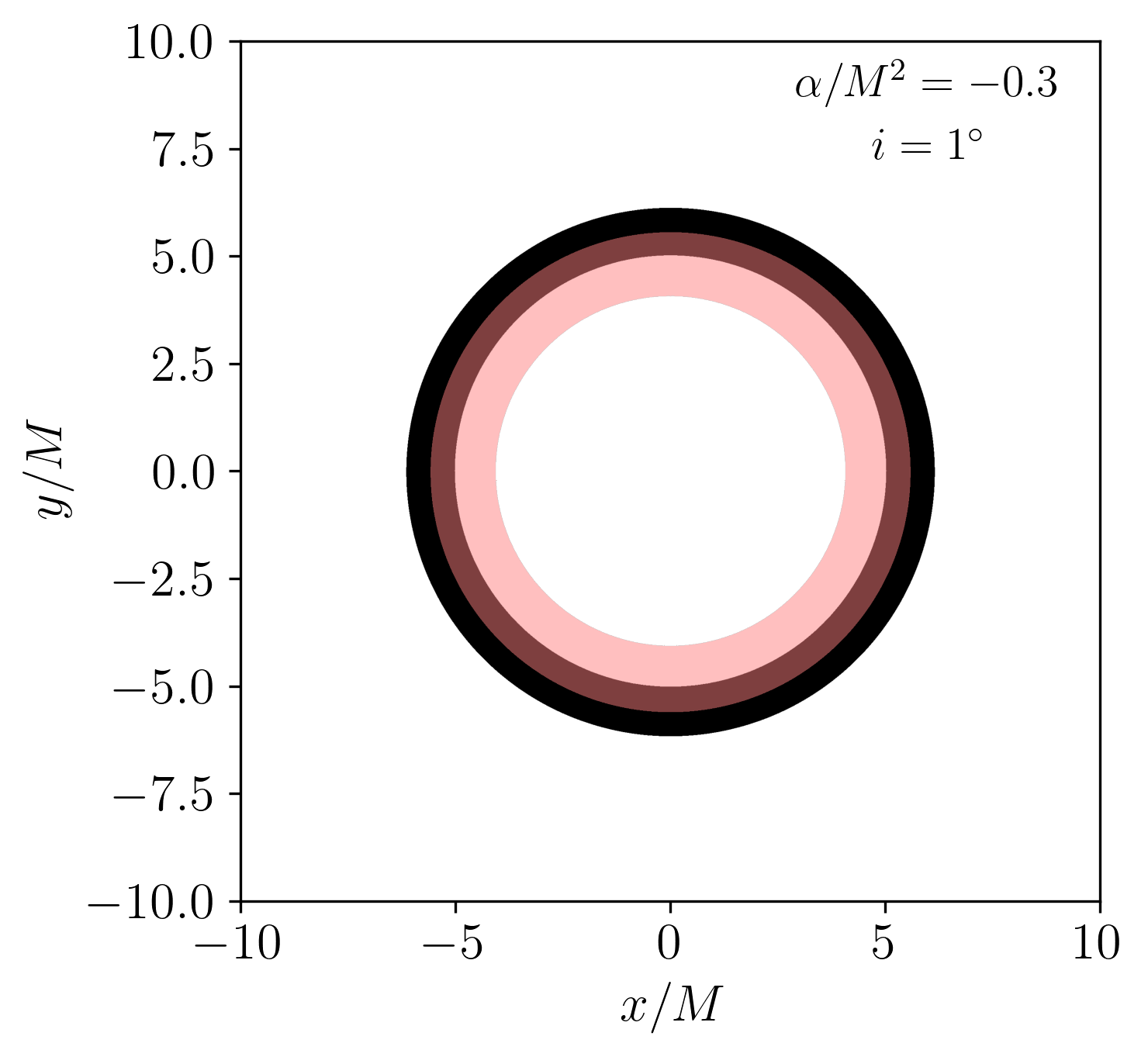}\hfill
    \includegraphics[width=0.333\linewidth]{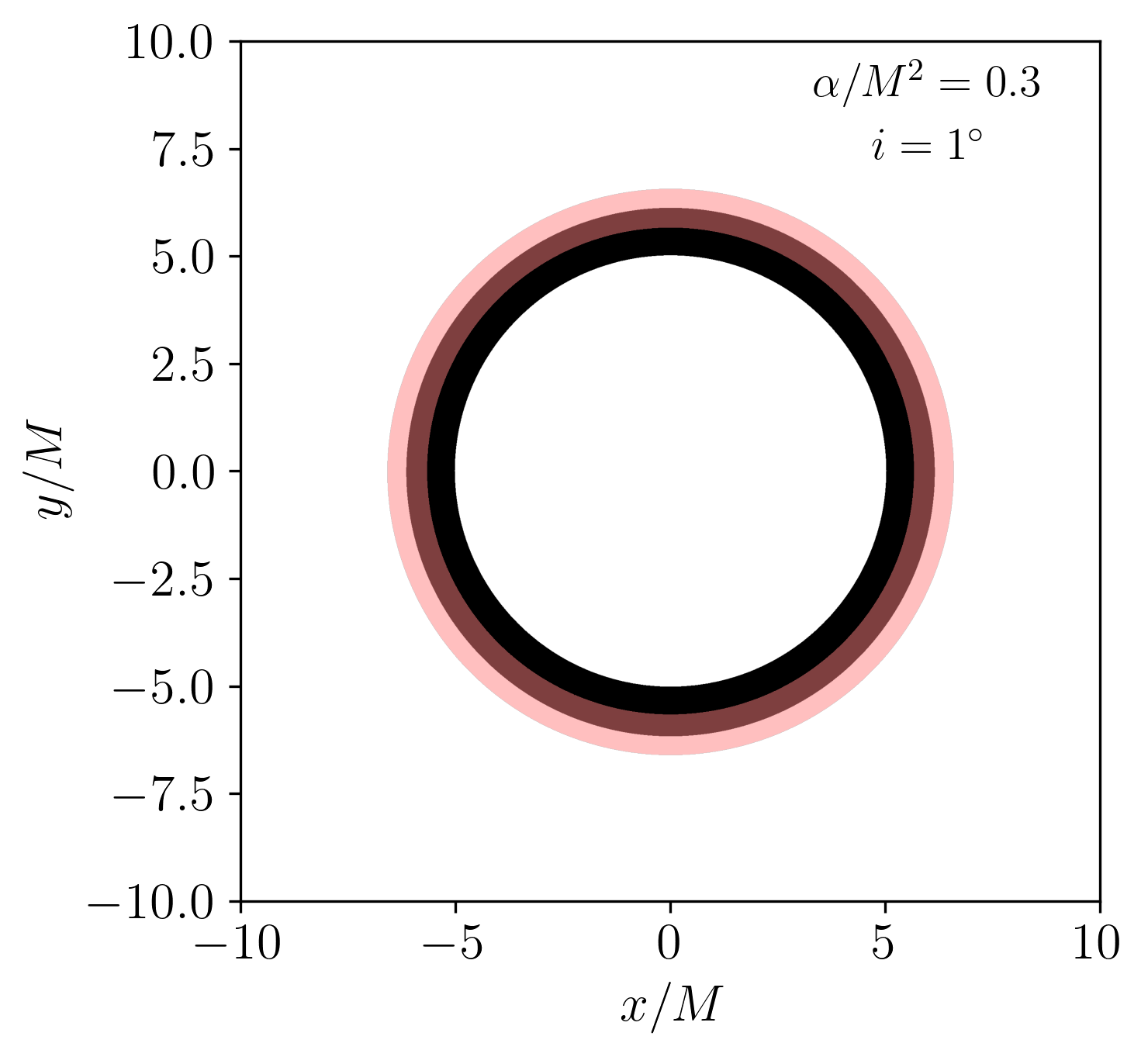}\hfill
    \includegraphics[width=0.333\linewidth]{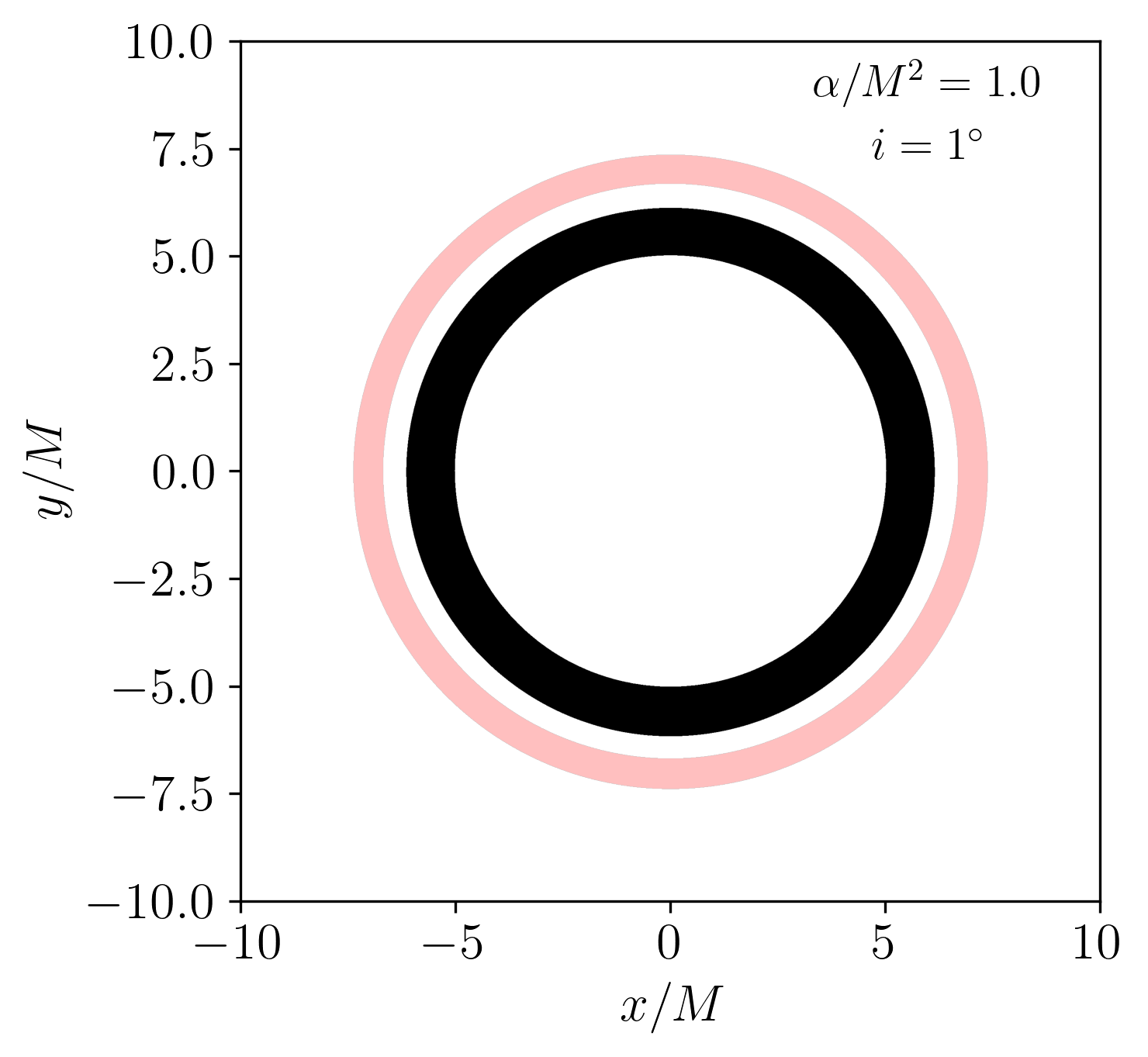}\vfill
    \includegraphics[width=0.333\linewidth]{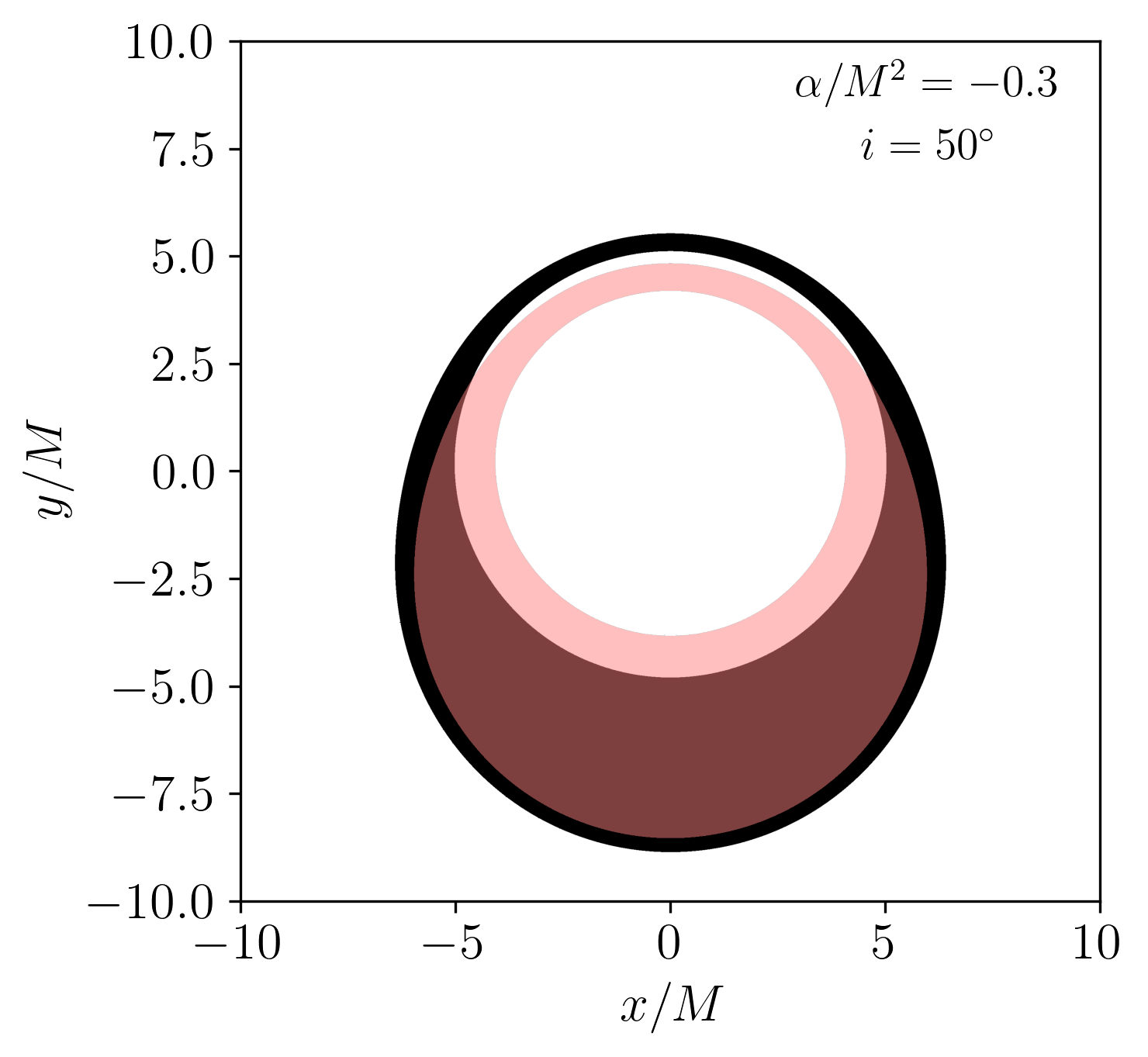}\hfill
    \includegraphics[width=0.333\linewidth]{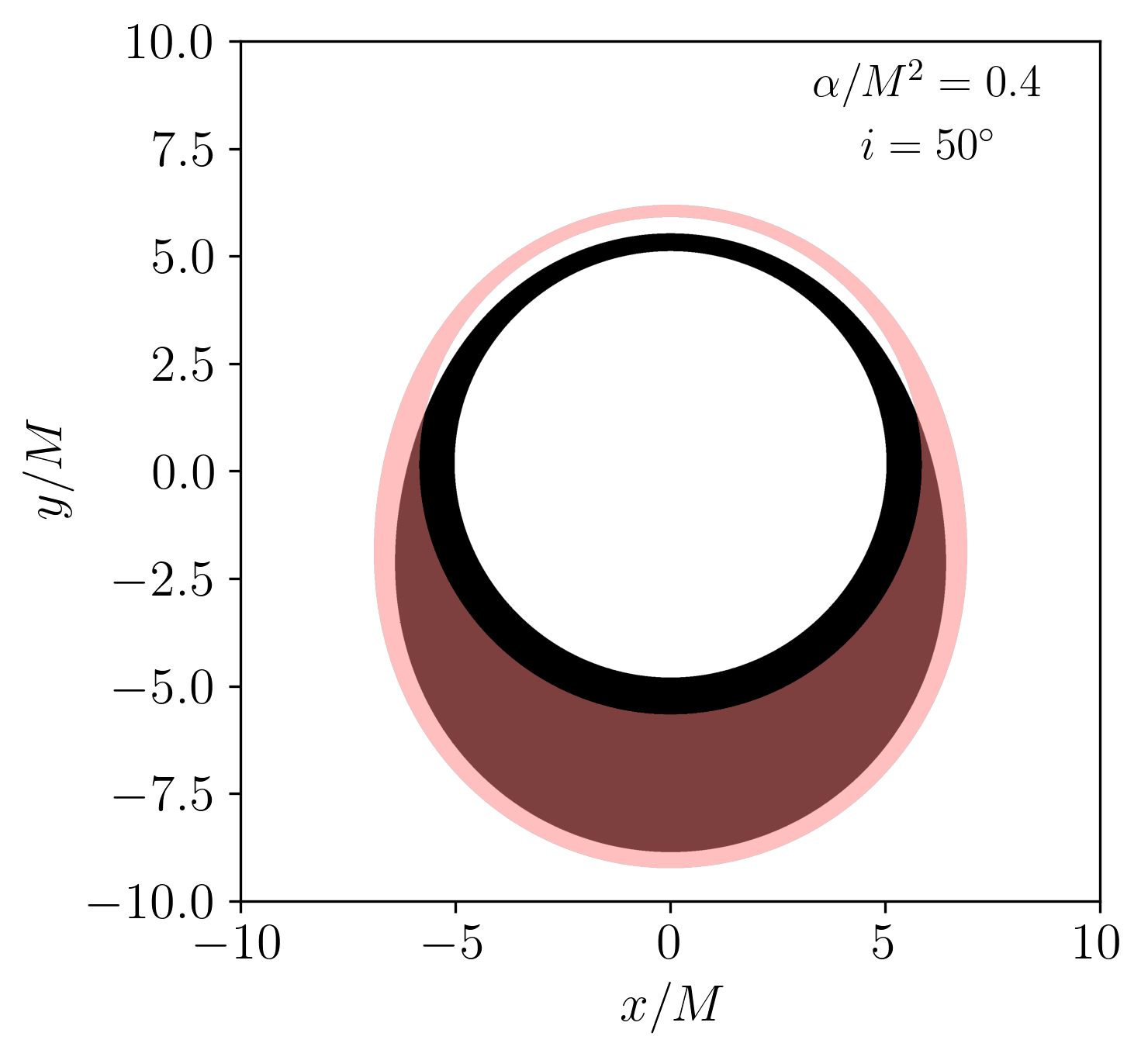}\hfill
    \includegraphics[width=0.333\linewidth]{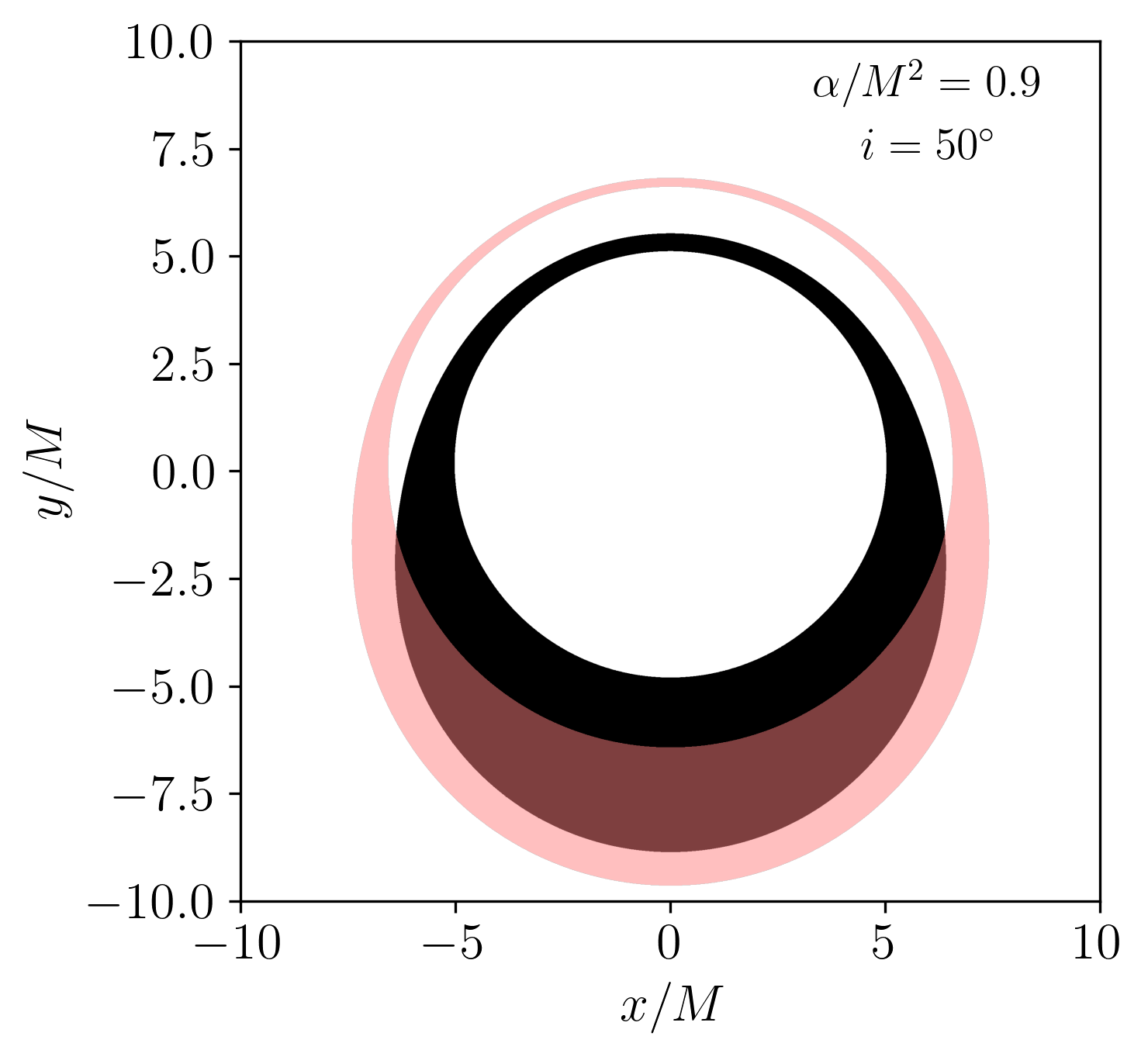}\vfill
    \includegraphics[width=0.333\linewidth]{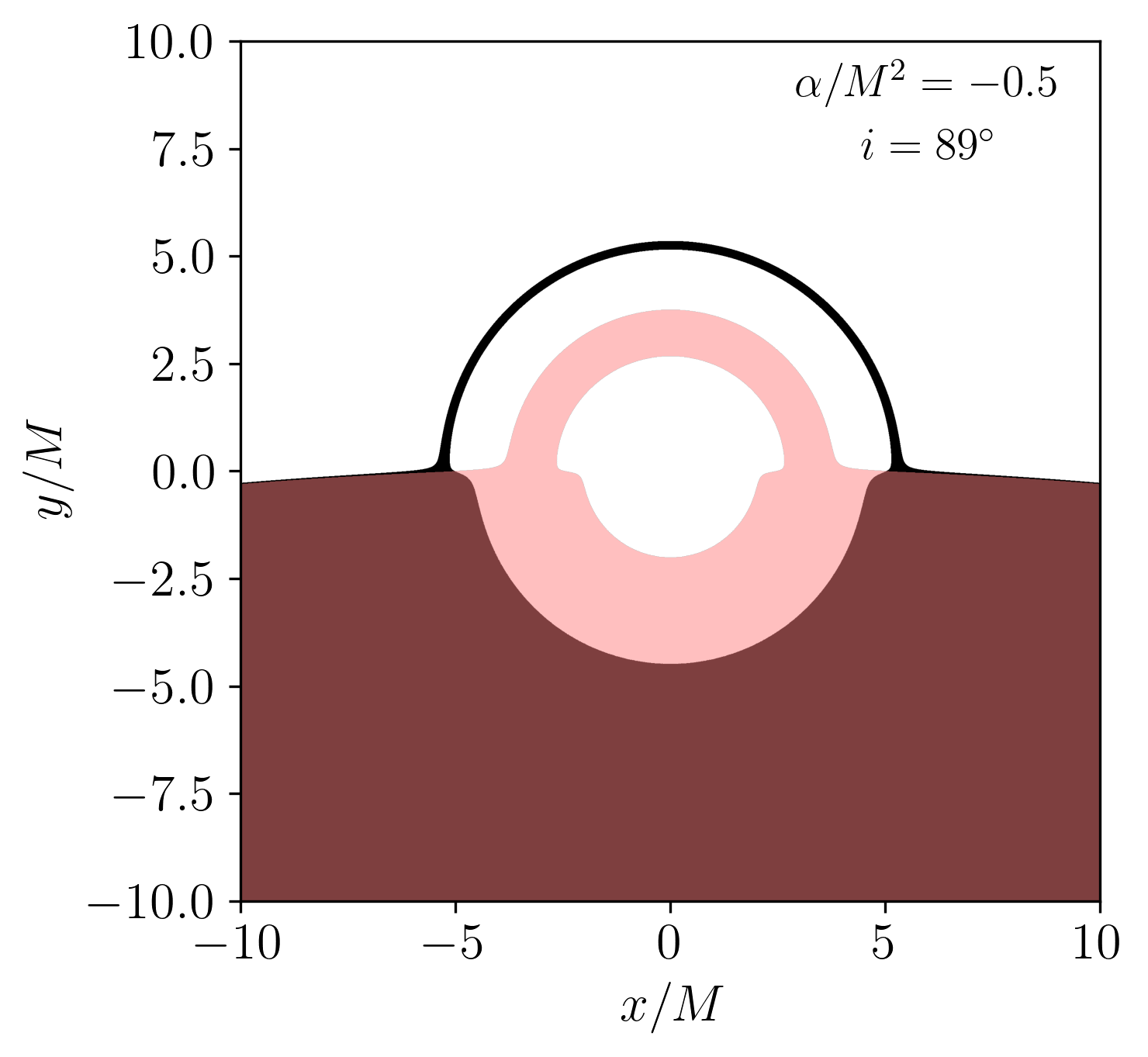}\hfill
    \includegraphics[width=0.333\linewidth]{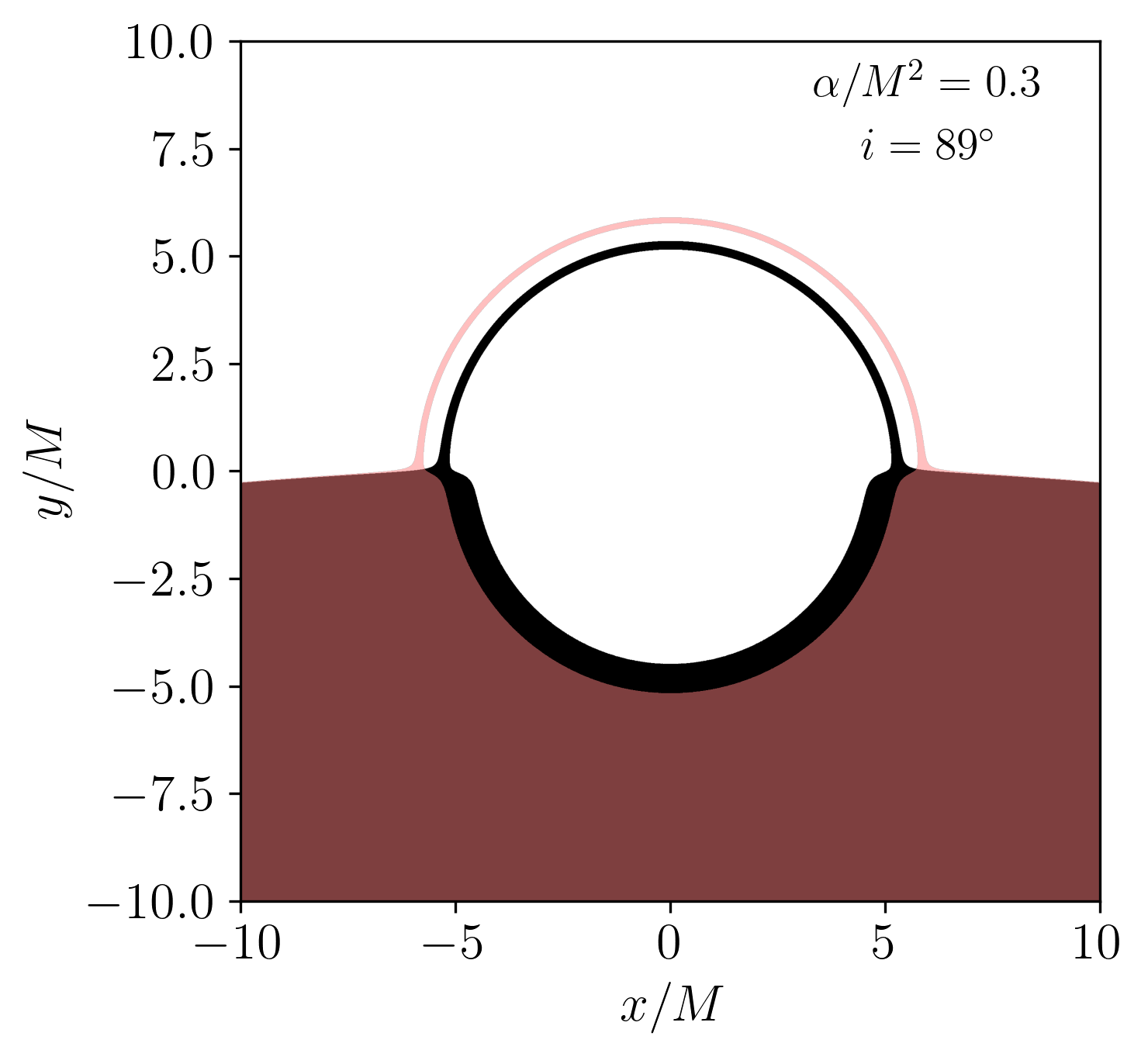}\hfill
    \includegraphics[width=0.333\linewidth]{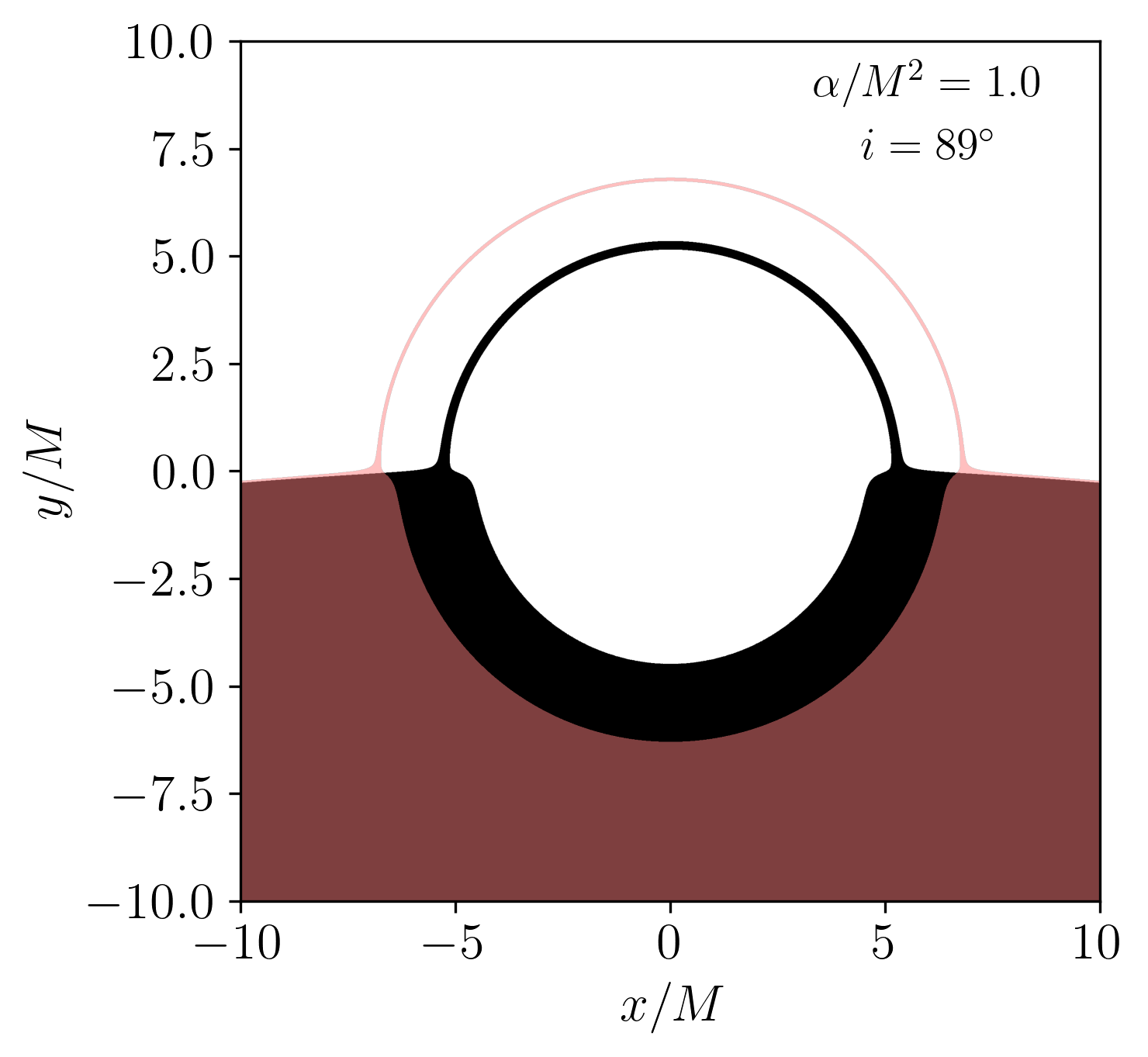}\vfill  
    \caption{We show the $n=1$ lensing band (which, for near edge-on inclinations, fills nearly the entire lower-half of the image) for the PPL polarization for three different inclinations $i$ and distinct coupling strengths $\alpha/M^2$. The lensing band for the Schwarzschild metric without non-minimal coupling is shown in black, and the lensing band with non-minimal coupling is overlaid in light pink. Wherever these regions overlap, the overlap is indicated in reddish brown. The lensing bands expand inwards (outwards) for negative (positive) value of the coupling. This behaviour is reversed for the PPM polarization.}
    \label{fig:PPL_lensing_bands_vs_Schwarzschild}
\end{figure*}
We calculate lensing bands with the Flexible Object Oriented Ray Tracer \texttt{FOORT}~\cite{Mayerson:2025xxx}, which ray-traces null geodesics in arbitrary spacetime geometries. \texttt{FOORT} has been previously used to explore image features of compact objects in modified theories of gravity~\cite{Mayerson:2023wck,Staelens:2023jgr,Fernandes:2024ztk}.
 
To obtain the lensing bands, we backwards ray-trace null geodesics in the background given by the effective metric for each polarization, and keep track of the number of half-orbits of each geodesic around the black hole. The results do not depend on the intensity of the emission, nor on assumptions about regions of higher absorptivity etc. This is because lensing bands dispense with all information regarding the intensity of light that arrives in a certain location in the image plane. Instead, lensing bands only encode all possible locations at which light arrives in the image plane after a given number of half-orbits about the black hole.
\begin{figure*}[t!]
	\centering
	\includegraphics[width=\linewidth]{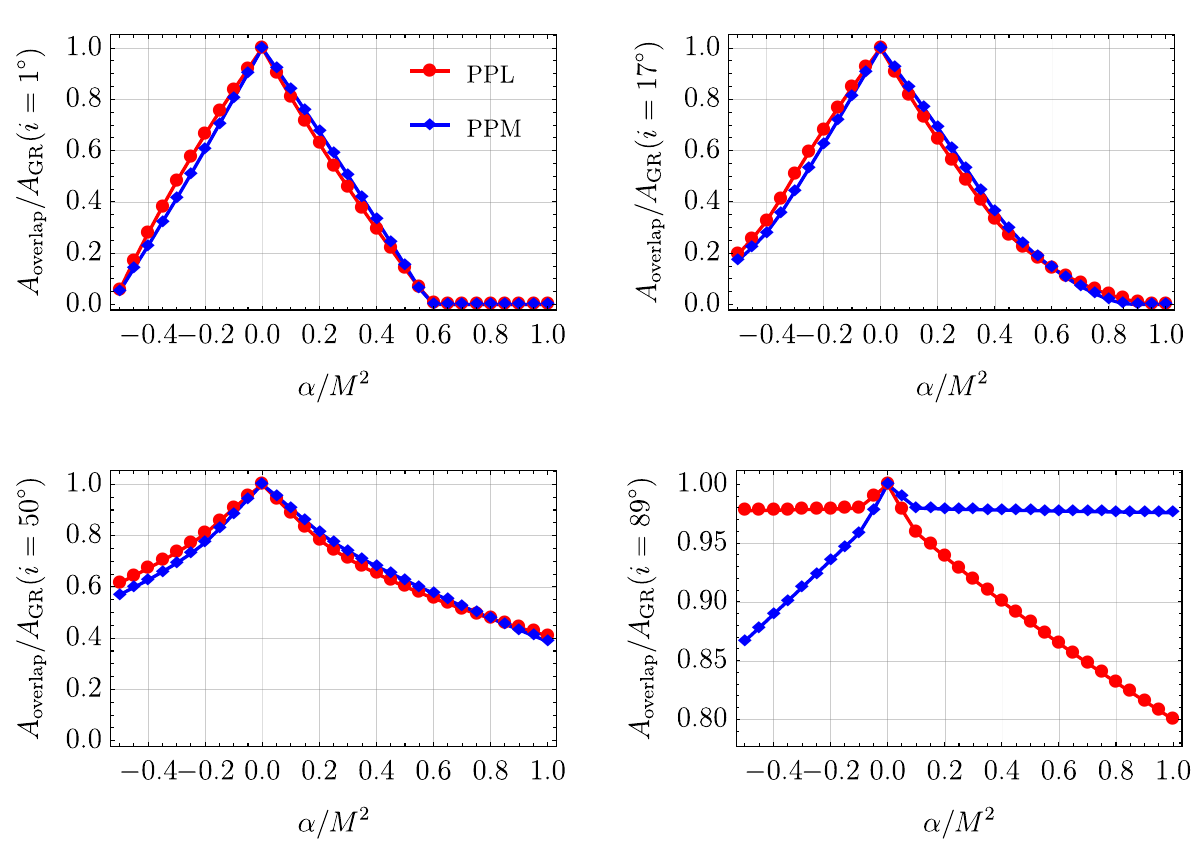}\vfill
    \caption{Area of the overlap between the $n=1$ lensing bands with non-minimal  coupling and GR (reddish brown regions in Fig.~\ref{fig:PPL_lensing_bands_vs_Schwarzschild} for the PPL case) in terms of the coupling strength $\alpha/M^2$, for four different inclinations $i$. The total field of view of the image is $20 M \times 20M$. For all inclinations, generically, the PPL lensing bands have a larger overlap with the GR lensing band, compared to the PPM case, for negative coupling, and vice-versa for positive couplings. For large inclinations, the overlap area can plateau because the lower-half of the image is filled by the lensing band both in GR and with non-minimal couplings.}
	\label{fig:plots_overlap}
\end{figure*}
We obtain lensing bands which depend on the ratio of the coupling $\alpha$ to the square of the black-hole mass $M$, as well as the inclination $i$, where $i=0 \degree$ corresponds to a line of sight between the observer and the black hole that is orthogonal to a would-be-accretion disk (or ``face-on''), and $i=90\degree$ corresponds to an ``edge-on" viewing angle.
We generate images for four distinct inclinations $i = \{1^\circ, 17^\circ, 50^\circ, 89^\circ\}$, with the observer placed at each inclination at a distance $r = 1000M$ from the black hole, using a field of view of size $20M \times 20M$. For all examined configurations, we produce high-resolution images with a minimum resolution of $512 \times 512$ pixels. Geodesics are traced  until they either cross the event horizon or reach a celestial sphere located at $r = 1000M$. For each inclination, we consider the whole range of couplings in Eq.~\eqref{eq:couplingrange}, starting from $\alpha/M^2 = -1/2$ in increments of $0.05$, until $\alpha/M^2 = 1$, such that for each polarization we have obtained a total of $120$ images with $\alpha/M^2 \neq 0$.

\section{Results}\label{sec:results}
\begin{figure*}[t!]
	\centering
	\includegraphics[width=\linewidth]{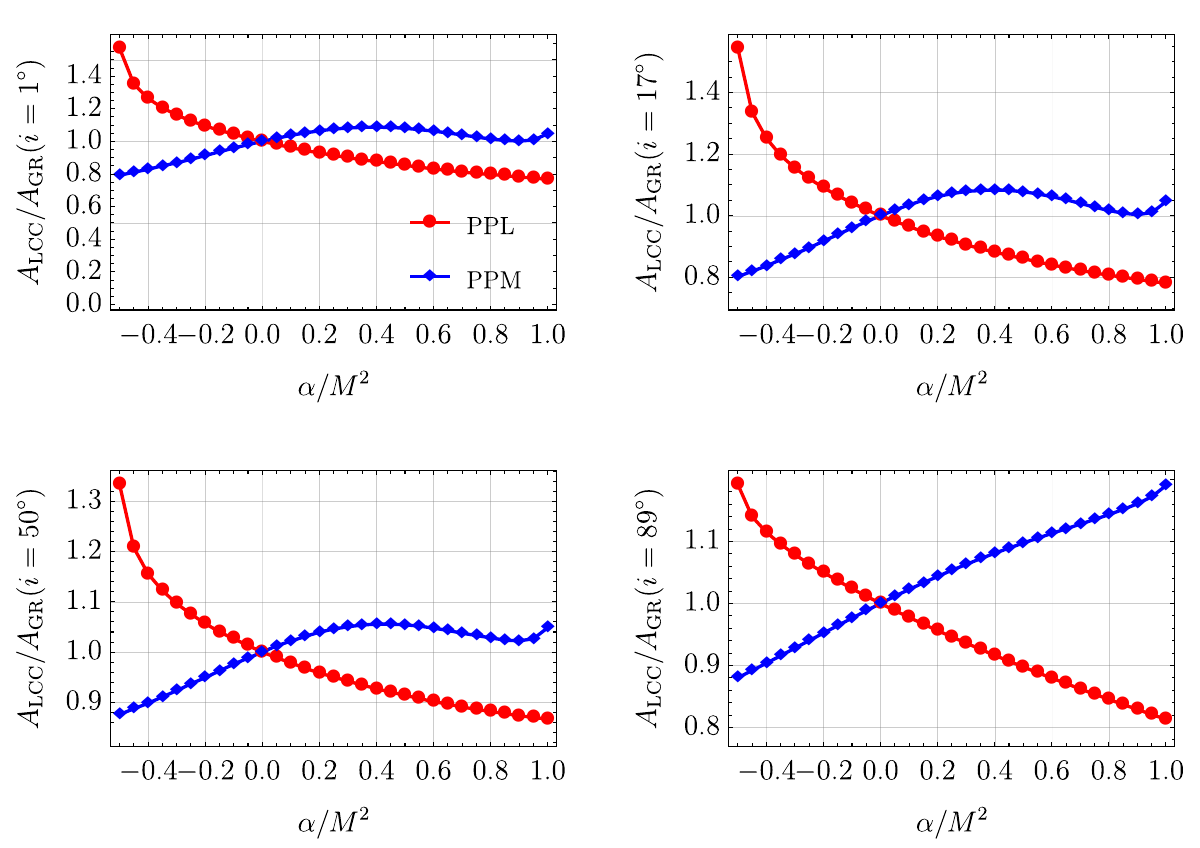}\vfill
    \caption{Ratio of the total area of the $n=1$ lensing bands with non-minimal light-curvature couplings (LCC) and GR in terms of the coupling strength $\alpha/M^2$ for four different inclinations $i$. This figure compares how much larger or smaller, compared to GR, is the region in the image plane where the $n=1$ photon ring can be located. The total area of the image is $20 M \times 20M$. For negative values of the coupling, there is a larger disparity between the two polarizations for small inclinations, making them more favorable to constrain the coupling.}
	\label{fig:plots_coupling_gr}
\end{figure*}
We compare the $n=1$ lensing bands of Schwarzschild black holes with and without non-minimal coupling in Fig.~\ref{fig:PPL_lensing_bands_vs_Schwarzschild}, for the PPL polarization. 

We observe that for negative coupling, the lensing band has a smaller diameter, whereas for positive coupling, it has a larger diameter. We also see that for $\alpha/M^2 \sim \mathcal{O}(1)$, the lensing bands with and without non-minimal coupling have no overlap. This sets the range of coupling values which we expect  a photon-ring-detection for M87* could robustly constrain. 

For inclinations $i \gg 0 \degree$, the overlap between lensing bands with and without non-minimal coupling depends on the angle in the image: for instance, for $\alpha/M^2 = 0.4$ and $i=50 \degree$, there is no overlap between lensing bands in the upper-half of the image, whereas there is an overlapping area in the lower part of the image. This is because, as the inclination increases, the upper part of the lensing band becomes thinner, whereas the lower part stretches out until it covers the entire lower-half of the image. 

To diagnose in a more quantitative way how large the overlap between lensing bands is and to find the critical values of the coupling (both positive and negative) when no overlap exists, we calculate the overlap area in the following, for both polarizations.

\subsection{Quantitative comparison of lensing bands to the case without non-minimal coupling}

In Fig.~\ref{fig:plots_overlap} we plot the overlap area between the $n=1$ lensing bands of the Schwarzschild black hole with and without the non-minimal coupling for both polarizations, normalized to the area of the $n=1$ lensing band of a Schwarzschild black hole without the non-minimal coupling (denoted as $\rm GR$ in the images). The two polarizations (PPL and PPM) strongly differ only at large inclination, i.e., close to edge-on. For both polarizations, the area of overlap with the $n=1$ Schwarzschild lensing band increases for $\alpha/M^2 < 0$, reaches a maximum at $\alpha/M^2 = 0$ (where the non-minimal coupling vanishes and we are back to GR) and decreases for $\alpha/M^2 > 0$ (except for $i = 89^{\circ}$ where PPL has a plateau for $\alpha/M^2 < 0$ and PPM a plateau for $\alpha/M^2 > 0$). For both polarizations, there are sufficiently large values of $\alpha/M^2$ for which there is no overlap: $\alpha/M^2 \gtrsim 0.6$ at $i = 1^{\circ}$ and $\alpha/M^2 \gtrsim 0.9$ at $i = 17^{\circ}$ (inclination of M87*). Those large values of $\alpha$ delineate a regime where the lensed photons for GR and for a non-minimal coupling are contained within fully distinct bands on the observer's screen. To exclude the corresponding values of couplings based on observations, it is necessary to combine EHT-observations with an independent mass measurement that sets the expected diameter of the photon ring in the image plane. Without such an independent mass measurement, a rescaling of the mass can always bring the photon ring diameter with non-minimal coupling into agreement with a photon ring without non-minimal coupling and at appropriately rescaled mass.

In Fig.~\ref{fig:plots_coupling_gr} we compare the absolute areas on the image plane of the $n=1$ lensing bands with non-minimal coupling with those of GR. At each inclination, there is a crossing between PPL and PPM at the GR value $\alpha/M^2 = 0$: PPL dominates for $\alpha/M^2 < 0$ and PPM takes over for $\alpha/M^2 > 0$. Hence, the area spanned by the lensing bands with non-minimal couplings is larger (smaller) than in GR as long as $\alpha/M^2 < 0$ and smaller (larger) than in GR for $\alpha/M^2 > 0$ for PPL (respectively, PPM). 

\begin{figure}[t!]
	\centering
	\includegraphics[width=\linewidth]{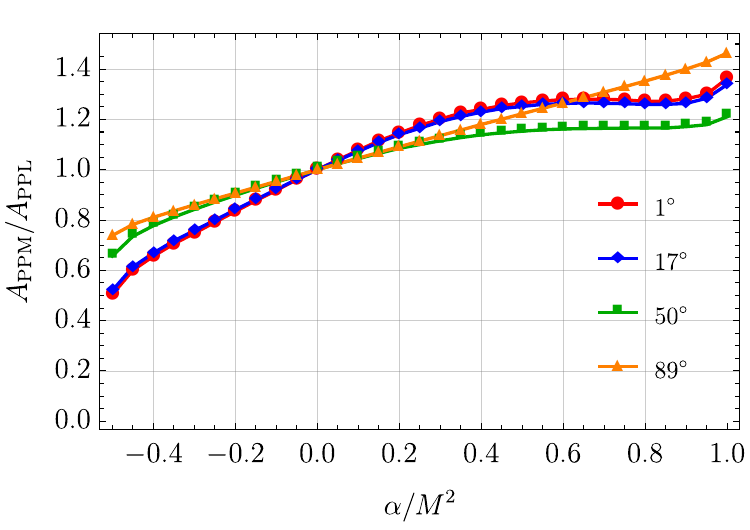}\hfill
	\includegraphics[width=\linewidth]{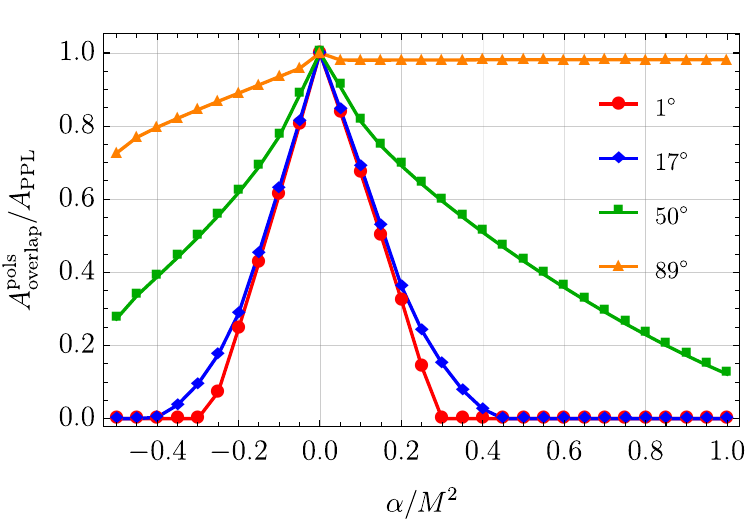}\vfill
    \caption{Top: Ratio of the area of the $n=1$ lensing bands for the PPM and PPL polarizations in terms of the coupling strength $\alpha/M^2$, for four different inclinations $i$. Bottom: Ratio of the area of the overlap between the $n=1$ lensing bands for the PPM and PPL polarizations and the area of the PPL $n=1$ lensing band in terms of the coupling strength $\alpha/M^2$, for four different inclinations $i$. The total area of the image is $20 M \times 20M$. For large inclinations, the overlap area plateaus because both polarizations fill a large portion of the lower-half plane of the image, and no overlap exists in the upper-half plane of the image. For sufficiently small inclinations, and large absolute values of the coupling, there is no overlap between the lensing bands, as increasing $|\alpha|$ pushes them in different directions.}
	\label{fig:plots_two_pols}
\end{figure}

In Fig.~\ref{fig:plots_two_pols} we compare the area of the $n=1$ lensing bands of the two distinct polarizations. We observe that $A_{\rm PPM} < A_{\rm PPL}$ for $\alpha/M^2 < 0$, while $A_{\rm PPM} > A_{\rm PPL}$ for $\alpha/M^2 > 0$, with very weak dependence in the inclination, in agreement with Fig.~\ref{fig:plots_coupling_gr}. More interestingly, for low-enough inclinations $i \lesssim 17^{\circ}$, there are intervals $-1/2 < \alpha/M^2 \leq -0.4$ and $0.4\leq \alpha/M^2 < 1$ for which there is no overlap between the lensing bands of the two polarizations. This is in strong contrast to the case without non-minimal coupling, where the $n=1$ lensing band has no polarization dependence. It also opens an intriguing observational possibility, because the EHT actually reports information on polarization, see, e.g., \cite{EventHorizonTelescope:2021bee,EventHorizonTelescope:2023gtd,EventHorizonTelescope:2024hpu,EventHorizonTelescope:2024rju}. Therefore, a reconstruction of the $n=1$ photon ring in two polarizations corresponding to PPL and PPM constrains $\alpha$ without the need for an independent measurement of the mass of the black hole.

\begin{figure}[h!]
	\centering
	\includegraphics[width=\linewidth]{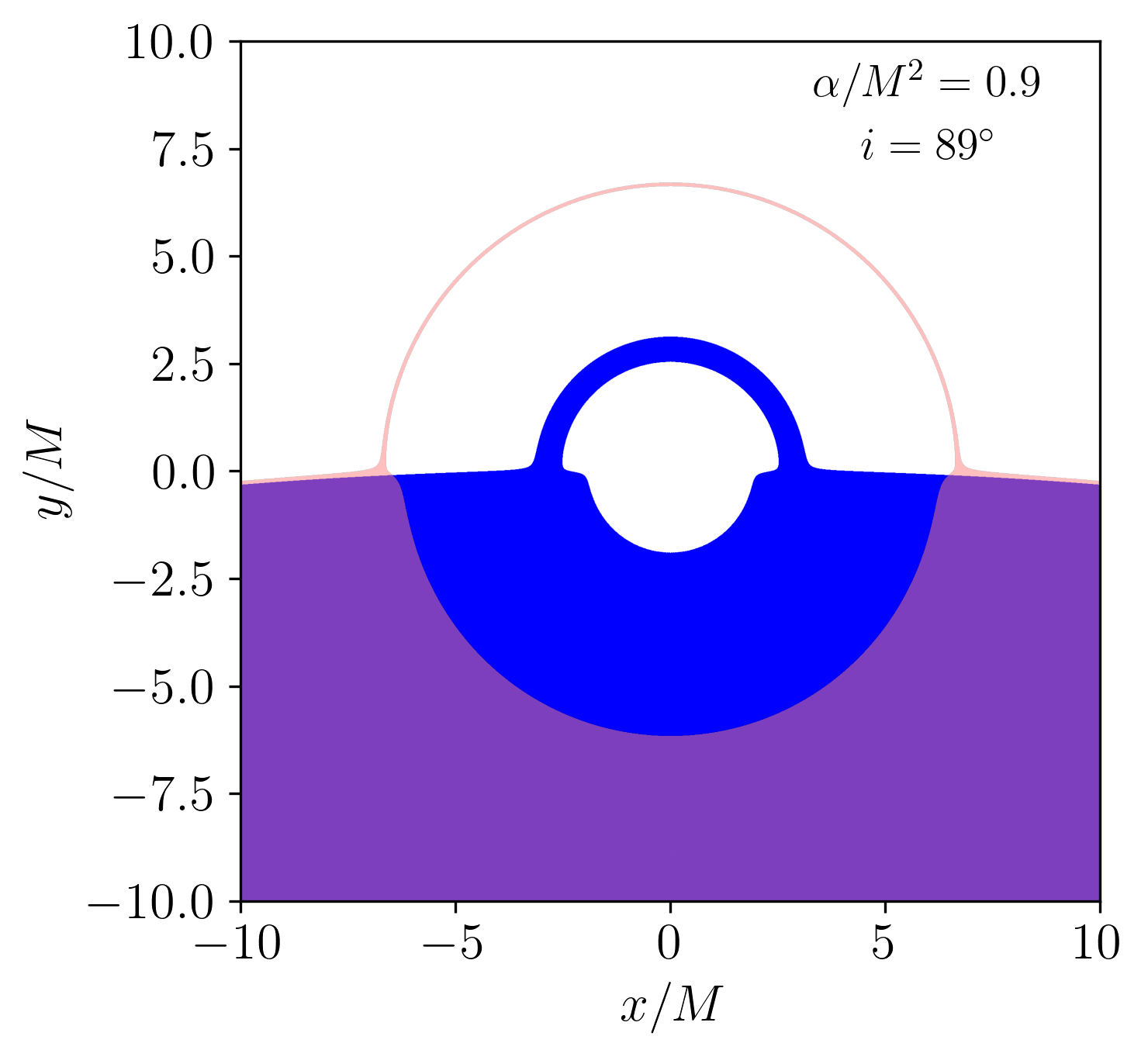}
    \caption{We show the $n=1$ lensing bands (which, for near edge-on inclinations, nearly fill the entire lower-half plane of the image) for the PPL and PPM polarizations for the largest inclination $i = 89^{\circ}$ and coupling strength $\alpha/M^2 = 0.9$. The total area of the image is $20 M \times 20M$. The lensing band for the PPL polarization is shown in light pink, and the lensing band for the PPM polarization is overlaid in blue. The region where these lensing bands overlap is indicated in purple. In the upper-half plane $y \geq 0$, the PPL lensing band is everywhere located outward from the PPM lensing band, while the PPL and PPM lensing bands greatly overlap in the lower-half plane $y < 0$, filling it almost entirely.}
	\label{fig:comp_plot_PPMvsPPL}
\end{figure}

\begin{figure}[h!]
	\centering
	\includegraphics[width=\linewidth]{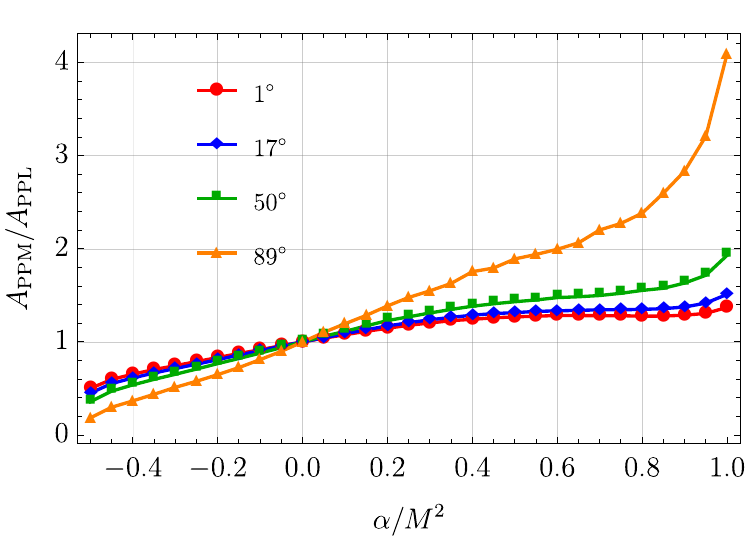}\hfill
	\includegraphics[width=\linewidth]{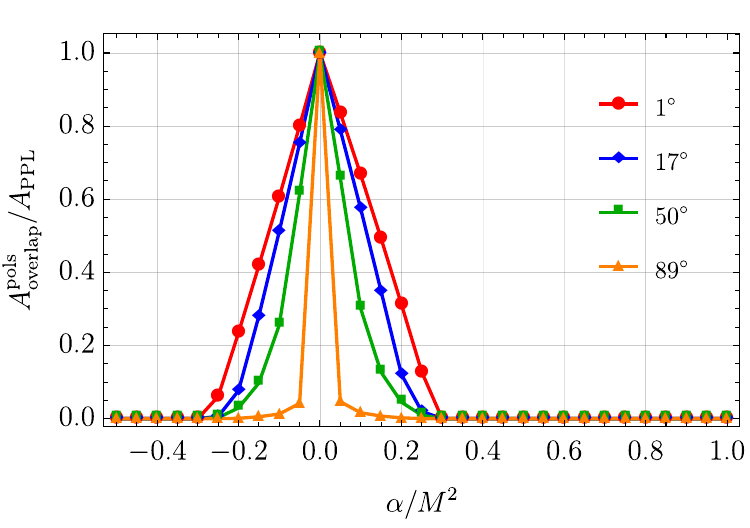}\vfill
    \caption{Top: Ratio of the area of the $n=1$ lensing bands for the PPM and PPL polarizations in terms of the coupling strength $\alpha/M^2$, for four different inclinations $i$, considering only the upper-half plane $y \geq 0$. Bottom: Ratio of the area of the overlap between the $n=1$ lensing bands for the PPM and PPL polarizations and the area of the PPL $n=1$ lensing band in terms of the coupling strength $\alpha/M^2$, for four different inclinations $i$, considering only the upper-half plane $y \geq 0$. The total area of the image is $10 M \times 20M$. The overlap area decreases with increasing inclination as the PPL and PPM lensing bands are clearly distinct in the upper-half plane at $i = 89^{\circ}$, see Fig.~\ref{fig:comp_plot_PPMvsPPL}.}
	\label{fig:plots_two_pols_upper_half}
\end{figure}

As one would expect, the area of overlap between the lensing bands of the two polarizations quickly grows as the GR case $\alpha/M^2 = 0$ is approached and only coupling values $\alpha \sim \mathcal{O}(M^2)$ can be constrained. Also, at high inclination $i = 89^{\circ}$, the overlap areas are similar for almost all values of $\alpha/M^2$ as $A_{\rm PPM} \sim A_{\rm PPL}$ due to the enlargement of the PPL and PPM lensing bands in the lower-half plane caused by relativistic effects. On the other hand, for high inclination angles there is no overlap in the upper-half plane of the image, as seen in Fig.~\ref{fig:comp_plot_PPMvsPPL}, and overall the lensing bands associated with both polarizations may remain distinguishable. These differences can be captured quantitatively by calculating the ratio between the lensing bands of the two polarizations restricted to the upper-half plane of the image, which we plot in Fig.~\ref{fig:plots_two_pols_upper_half}. We caution that actual observational data, with sparse coverage in the Fourier plane, may be much more difficult to interpret in terms of such observables. Nevertheless, our results indicate that it may be relevant to develop observables that are adapted to only parts of the resulting image, such as, in our case, only the upper half plane. 
We believe this is worth exploring in detail elsewhere.


\subsection{A case study: the M87* photon ring}
By imposing an informed prior on the image reconstruction of M87*, Ref.~\cite{Broderick:2022tfu} demonstrated that a persistent thin ring, visible across multiple observational days, is statistically preferred alongside a broader emission region when compared to reconstructions lacking such a ring. Thin photon rings generically appear in shadow images of both GR and beyond-GR (ultra-)compact objects~\cite{Guerrero:2021ues,Olmo:2021piq,Eichhorn:2022fcl,Carballo-Rubio:2022aed,Guerrero:2022msp,daSilva:2023jxa,Carballo-Rubio:2025fnc},  motivating this choice of prior. While the reconstructed thin ring may be interpreted as the $n=1$ photon ring, its presence ultimately depends on the adopted prior. As a case study to explore whether future VLBI-arrays can, in principle, constrain the non-minimal coupling we consider in this work, we take the results of Ref.~\cite{Broderick:2022tfu} at face value. That is, we assume that the thin ring matches  the $n = 1$ photon ring of M87* which, due to the low inclination $i_{\rm M87^*} \approx 17^\circ$ of M87*, is well-approximated by a circle. 

\begin{figure}[h!]
	\centering
	\includegraphics[width=\linewidth]{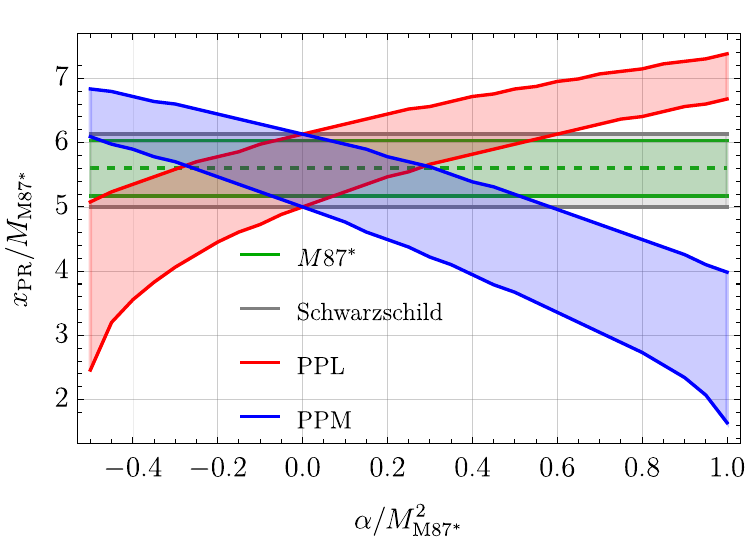}\vfill
    \caption{In the image plane, we fix $y=0$ (see the lensing bands in Fig.~\ref{fig:PPL_lensing_bands_vs_Schwarzschild}) and compare the allowed range of $x_{\rm PR}$ where the $n=1$ photon ring can be located (i.e., the lensing band) for the cases of Schwarzschild (gray), PPL (red) and PPM (blue) with the measured value (in green), assuming an inclination of $i = 17^\circ = i_{\rm M87^*}$. Regions where there is no overlap between the PPL and PPM cases are excluded as they are not compatible with unpolarized observations.}
	\label{fig:plot_pr}
\end{figure}

In this case, the angular radius of the M87* photon ring is $\theta_{\rm ring} = 21.74 \pm 0.10\, \mu{\rm as}$ \cite{Broderick:2022tfu}, the distance to M87* is $d=16.8\pm0.8\, {\rm Mpc}$ \cite{EventHorizonTelescope:2019ggy}, while external mass measurements from gas dynamics result in $M_{\rm M87^*} = (6.60 \pm 0.40) \cdot 10^9 {\rm M_{\odot}}$ \cite{Gebhardt:2011yw}. Using these values, we can obtain the dimensionless diameter $x_{\rm PR}/M_{\rm M87^*}$ of the photon ring in image-plane coordinates via
\begin{equation}
    \frac{x_{\rm PR}}{M_{\rm M87^*}} = \frac{\theta_{\rm ring} d}{M_{\rm M87^*}} \approx 5.60 \pm 0.43.
    \label{eq:PRM87}
\end{equation}
At the inclination $i=17^\circ$ and for the whole range of couplings, we take a cross section of our images at $y=0$ and compare the region along the $x$-axis where the $n=1$ photon ring  with non-minimal coupling can be located with the value for M87* given in Eq.~\eqref{eq:PRM87}. The results are shown in Fig.~\ref{fig:plot_pr}, from which we see that the only values of $\alpha/M^2_{\rm M87^*}$ compatible with unpolarized observations -- for which PPL and PPM lensing bands overlap along the $x$-axis -- are $-0.3 \lesssim \alpha/M^2_{\rm M87^*} \lesssim 0.3$.\footnote{Most of the analyses of the bright thin ring apparent in shadow images of M87* have been performed by the EHT Collaboration with unpolarized light, as the linearly- and circularly-polarized components of the lensed radiation captured by the current VLBI-array are found to be subdominant, but have nevertheless been successfully extracted \cite{EventHorizonTelescope:2021bee,EventHorizonTelescope:2021srq,EventHorizonTelescope:2023gtd,EventHorizonTelescope:2025vum}.} These constraints result in the upper bound
\begin{equation}
    \sqrt{|\alpha|} \lesssim 5.34 \times 10^{9} \,\mathrm{km}.
    \label{eq:constraint}
\end{equation}
\section{Conclusions}\label{sec:conclusions}
In this work we have explored non-minimal light-curvature couplings and how they affect black-hole images, in particular the $n=1$ lensing band of the Schwarzschild black hole. Following Ref.~\cite{Drummond:1979pp}, see also \cite{Chen:2015cpa}, we used that non-minimally coupled photons propagate along geodesics of effective, polarization-dependent metrics. This description holds within a finite range of values of the non-minimal coupling $\alpha$.

We are motivated by the capabilities of VLBI arrays to constrain deviations from the Kerr (or more specifically Schwarzschild) spacetime, which has led to a large body of literature~\cite{Broderick:2013rlq,Wei:2013kza,Grenzebach:2014fha,Johannsen:2015hib,Cunha:2015yba,Cunha:2016wzk,Vincent:2016sjq,Abdujabbarov:2016hnw,Amir:2016cen,Tsukamoto:2017fxq,Ayzenberg:2018jip,Psaltis:2018xkc,Mizuno:2018lxz,Cunha:2018acu,Held:2019xde,Tian:2019yhn,Vagnozzi:2019apd,Allahyari:2019jqz,Cunha:2019dwb,Volkel:2020xlc,Konoplya:2020bxa,Khodadi:2021gbc,Younsi:2021dxe,EventHorizonTelescope:2022xqj,Vagnozzi:2022moj,Eichhorn:2022fcl,Carballo-Rubio:2022aed,Chen:2022scf,Khodadi:2020jij,Fernandes:2024ztk,dePaula:2023ozi,Sengo:2022jif,Junior:2021dyw,Xavier:2020egv,Olmo:2023lil,Ayzenberg:2023hfw,Carballo-Rubio:2023ekp,Afrin:2022ztr,Salehi:2023eqy,Lupsasca:2024wkp,Broderick:2024vjp,Khodadi:2024ubi,Carballo-Rubio:2025fnc,Bambi:2025wjx}. Constraints are placed on deviation parameters in the spacetime metric or couplings in Lagrangians. These constraints bound dimensionless ratios of deviation parameters/couplings to the gravitational radius which are of $\mathcal{O}(1)$. In a standard EFT setting, one expects that the same scale suppresses all higher-order interactions. Thus, a treatment consistent with this expectation also has to account for non-minimal couplings. The specific coupling we study here is already constrained from solar-system tests and neutron-star observations. If one takes these constraints at face value\footnote{In other astrophysical environments, such as, e.g., around neutron stars, other non-minimal couplings, e.g., $R F_{\mu\nu} F^{\mu\nu}$, may also play a role. The translation of observations into bounds on couplings is therefore subtle and depends on assumptions about which couplings are present in the Lagrangian; i.e., the astrophysical constraints are model-dependent in that sense.}, the coupling values we consider are already ruled out for supermassive black holes.\footnote{We would like to point out that supermassive black holes may become interesting as a testing ground for such couplings, thanks to the newly proposed method of intensity interferometry \cite{Dalal:2024aaj} which, for an observing frequency of $550$ THz and baselines of order $10^4$ km, could resolve angular scales up to $\theta \sim 0.01\,\mu$as. This would correspond to an improvement in resolution by a factor $2000$ compared to the current EHT resolution and match the range of angular scales associated with the $n=1$ photon ring of supermassive black holes.}  However, we consider this coupling as a paradigmatic example, which creates effects in black-hole lensing bands which are likely to generalize across many other models with non-minimal couplings. Our main goal is therefore explicitly not to provide competitive bounds on $\alpha$ or to identify the best possible environment to constrain $\alpha$. Rather, we aim at preparing the ground for future studies of new physics in gravity, which we, in general, expect to affect both spacetime geometry as well as propagation of light. Understanding the effects that come from modified propagation is therefore crucial. The specific coupling $\alpha$ serves as a useful example of a more general class of couplings.

We compared the $n=1$ lensing band of a Schwarzschild black hole with and without non-minimal coupling $\alpha$. For $\alpha/M^2 \sim \mathcal{O}(1)$, the resulting deformations of lensing bands can become large for both photon polarizations. We find the following two results: First, in the most extreme cases within the regime of validity of the effective-metric description, the $n=1$ lensing bands with and without non-minimal coupling do not overlap. Second, lensing bands of the two polarizations differ from each other.
From these results, we conclude the following:
\begin{enumerate}
\item[a)] Qualitatively, non-minimal couplings and deformations of the spacetime geometry can result in similar observational signatures, namely deformed lensing bands. There is thus a possibility that those effects are degenerate and/or modifications of the spacetime geometry and non-minimal couplings may compensate to produce GR-like images, even though the underlying theory is very different from GR.
\item[b)] Given an external mass measurement, e.g., from stellar orbits \cite{Gebhardt:2011yw} (and a spin measurement in the more realistic Kerr case), which fixes the expected photon-ring diameter in GR without non-minimal couplings, and a measurement of the photon-ring diameter, we can rule out a range of values of $\alpha$, under the assumption that the underlying spacetime geometry is that of GR. 
\item[c)] Because the lensing band in GR with minimal light-curvature couplings does not depend on the polarization, a measurement of the photon-ring diameter in two different polarizations leads to constraints on the non-minimal coupling without the need for an external mass-measurement.
\item[d)] Because modifications of the spacetime geometry in a given theory may not lead to different lensing bands for different polarizations, it may be possible to disentangle some modifications of the geometry from non-minimal couplings.
\end{enumerate}

As an avenue for future research, it would be interesting to generalize our results to rotating black holes, since astrophysical supermassive  black holes are expected to have non-negligible spin. The effective metrics that photons of different polarizations follow in this case have been derived in Ref.~\cite{Chen:2023wna}. Interestingly, those effective metrics break circularity symmetry \cite{Papapetrou:1966zz,Carter:1973rla,Carter:2009nex}, which may occur
for black-hole spacetime metrics in scenarios beyond GR \cite{Delaporte:2022acp,Babichev:2025szb} like modified gravity and quantum gravity \cite{Minamitsuji:2020jvf,Anson:2020trg,BenAchour:2020fgy,Adam:2021vsk,Eichhorn:2021iwq,Fernandes:2023vux}.

In addition, combining effects of non-minimal light-curvature couplings with modifications of the spacetime geometry will enable us to study whether these effects can be disentangled from each other.

Finally, going beyond the Horndeski coupling we considered here, there are additional non-minimal couplings not yet constrained observationally, which may have similar effects. These couplings are, in general, not yet constrained by any observations; therefore, constraints from observations of supermassive black holes may even be of quantitative interest.

Ultimately, measurements of black-hole photon rings, as proposed in \cite{Johnson:2024ttr,Lupsasca:2024xhq}, may result in constraints on various non-minimal couplings. Because these are also constrained (in combination with higher-order field-strength couplings) from positivity bounds, such constraints may provide indirect information on the properties of a more fundamental theory of gravity and matter. \\

\section*{Acknowledgments}
\noindent We gratefully acknowledge helpful discussions with Aaron Held.

This work has been supported by a grant from Villum Fonden under Grant No.~29405. This work is funded by the Deutsche Forschungsgemeinschaft (DFG, German Research Foundation) under Germany’s Excellence Strategy EXC 2181/1 - 390900948 (the Heidelberg STRUCTURES Excellence Cluster). RCR acknowledges financial support provided by the Spanish Government through the Ram\'on y Cajal program (contract RYC2023-045894-I), the Grant No.~PID2023-149018NB-C43 and the Severo Ochoa grant 
CEX2021-001131-S funded by MCIN/AEI/ 10.13039/501100011033, and by the Junta de Andaluc\'{\i}a 
through the project FQM219, as well as the hospitality of the Center of Gravity, a Center of Excellence funded by the Danish National Research Foundation under grant No.~184. 

\bibliographystyle{utphys}
\bibliography{refs}

\end{document}